\documentstyle[pra,aps,epsfig, multicol]{revtex}
\begin{document}
\draft
\title{
Collective excitations of
a two-dimensional interacting Bose gas in
anti-trap and  linear external potentials}
\author{D. V. Fil\footnote{e-mail fil@isc.kharkov.com}}
\address{
Institute for Single Crystals, National Academy of Sciences of
Ukraine, Lenin av. 60, Kharkov 61001, Ukraine}
\author{S. I. Shevchenko}
\address{
B. I. Verkin Institute for Low Temperature Physics and
Engineering, National Academy of Sciences of
Ukraine, Lenin av. 47, Kharkov 61164, Ukraine}
\maketitle
\begin{abstract}
We present a method of finding approximate analytical solutions for
the spectra and eigenvectors of collective modes in
a two-dimensional system of interacting bosons subjected to a
linear external potential or
the potential of a special form
$u(x,y)=\mu -u \cosh^2 x/l$, where $\mu$ is the chemical potential.
The eigenvalue problem is solved analytically for an artificial model
allowing the unbounded  density of the particles. The spectra of collective
modes are calculated numerically for the stripe,  the rare density
valley and the edge geometry and compared with the analytical results.
It is shown that the energies of the modes localized at the rare density
region and at the edge are well approximated by the analytical expressions.
We discuss Bose-Einstein condensation (BEC) in the systems under
investigations  at $T\ne 0$ and
find that in case of a finite number of the  particles the regime of
BEC can be realized, whereas the condensate disappears in the
thermodynamic limit.
\end{abstract}

\section{INTRODUCTION}
The observation of BEC in alkali metal vapors \cite{1,2,3} confined
in a trap has opened the opportunities to investigate experimentally
this macroscopic quantum phenomenon.
That outstanding discovery  has
stimulated  theoretical studies  of BEC in
inhomogeneous Bose gases subjected to external fields
(see \cite{4} for review).
One of the problems under extensive investigation is
the influence of an external potential on properties of
low-dimensional Bose systems. It is known that
an external potential may cause BEC in 2D ideal Bose gases
at nonzero temperatures
\cite{5,6,7}. This problem was discussed in detail by
Mullin \cite{8}. Mullin has presented
rigorous arguments why
the Hohenberg theorem \cite{9}
ruling out BEC in 2D does not apply to the
systems considered in \cite{5,6,7}.

Recently, the possibility of BEC in interacting 2D Bose gases has
been investigated. Petrov, Holzmann and Shlyapnikov \cite{10} considered
a 2D Bose system confined in a harmonic oscillator potential. They
have concluded that two regimes can be realized: a true condensate
and a quasicondensate with fluctuating phase.
In the mean field approximation the properties of such a system have
been studied by Bayindir and Tanatar \cite{11}.
They demonstrate the similarity in thermodynamic
behavior of ideal 2D Bose systems in  a harmonic trap
and those with weak interactions and a finite number of the particles.
On the other hand, Mullin \cite{8,12} argues the absence of BEC in
trapped interacting Bose gases in 2D in the thermodynamic limit.

For understanding the macroscopic quantum
properties of interacting Bose systems the key point
is the behavior of elementary excitations.
In two dimensions thermally excited fluctuations
destruct the condensate.
Therefore, the information about stability of
BEC can be gained from
the study of the spectrum and structure of low-energy excitations.

The theory for elementary excitations of interacting
Bose systems has been derived by Bogoliubov \cite{13}.
After the experiments \cite{1,2,3} this theory
has received a new interest.
Stringari \cite{14} has obtained an analytical solution
for the collective mode spectrum of a 3D Bose gas in an
isotropic harmonic trap in the Tomas-Fermi regime.
The same problem for the anisotropic trap has been resolved in
Ref.\ \cite{15}.  Numerical solutions of the Bogoliubov equations
for the trap geometry have been
obtained in Ref.\ \cite{16,17,18,19,20}.
The exact result for the energies of the harmonic trap
breathing modes in 2D has been found by Pitaevskii and Rosch \cite{21}.
The collective mode spectrum in a spatially periodic potential has been
investigated in Ref.\ \cite{22}.

Further understanding of the properties
of inhomogeneous 2D Bose systems requires the study of the
low-energy excitations
in the potentials different from the mentioned above. It is
important to investigate the potentials, for which
an analytical solution of the eigenvalue problem can be
found. In this paper we analyze two such cases.
The first case corresponds to the potential
$u(x,y)=\mu -u \cosh^2 x/l$, where $\mu$ is the chemical potential.
Since  the external field  pushes the particles away from the region $x=0$,
and a rare density valley is formed at  $x=0$, we call such a geometry
an "anti-trap" one.
The second potential considered is a linear one:
$u(x,y)=\mu -\alpha x$. The Bose cloud subjected to that potential
has an edge at $x=0$
with a linear dependence of the density on the distance
to the edge.

An approximate analytical solution of the eigenvalue problem
is obtained by the following method.
We formally imply the density is unbounded at
$|x|\to\infty$ ($x\to +\infty$ for the linear potential).
Then, analytical expressions for the energies and the
eigenvectors of collective excitations  can be derived.
To establish the applicability of the model description to real
objects the eigenvalue problem is also solved numerically.
It is shown that the spectra of collective modes localized at the
rare density region (in the anti-trap potential) and at the edge (in the
linear potential) are well approximated by the analytical solution.
Using the analytical solution we calculate the one particle
density matrix and analyze the
off-diagonal long range order. We conclude that
thermal fluctuations
do not destroy completely
the long range order in the systems with a finite number of the particles.
It means that the BEC regime is realized.
It is not the case in the thermodynamic limit, where the density of the
condensate approaches to zero at $T\ne 0$.

In Sec.\ \ref{s2}
a general description of elementary excitations in
inhomogeneous 2D Bose systems is presented.
The collective mode spectra in the anti-trap and the linear potential
are obtained in Sec.\ \ref{s3} and Sec.\ \ref{s4}, correspondingly.
The one particle density matrix is calculated in Sec.\ \ref{s5}, and
the  conclusions are given in Sec.\ \ref{s6}.

\section{COLLECTIVE EXCITATIONS OF INHOMOGENEOUS 2D BOSE GAS}
\label{s2}
We consider an interacting 2D Bose gas in the external potential
$u({\bf r})$. The Hamiltonian of the system can be written in the
following form
\begin{equation}
H=\int d^2 r \{{\hbar^2\over 2 m} \nabla \
\hat{\psi}^{+}({\bf r}) \nabla \hat{\psi} ({\bf r})+
[u({\bf r})-\mu ]\hat{\psi}^{+}({\bf r}) \hat{\psi}({\bf r})  +
{1\over 2}
 \gamma  \hat{\psi}^{+}({\bf r}) \hat{\psi}^{+}({\bf r})
\hat{\psi} ({\bf r}) \hat{\psi}({\bf r})\},
\label{1}
\end{equation}
where  $\hat{\psi}$ is the boson field
operator, $m$, the boson mass.
In Eq.\ (\ref{1}) we use the
effective potential of the atom-atom interaction
of the form
$V({\bf r}-{\bf r}')=\gamma \delta({\bf r}-{\bf r}')$.
The strength $\gamma$ is considered as
a parameter.
To describe the collective excitations
we use an approach
developed by one of the authors in Refs.\ \cite{23,24,25}.
We introduce the
phase  $\hat{\varphi}({\bf r})$ and
density $\hat{\rho}({\bf r})$ operators.
They are connected with
the Bose field operator by the relation
\begin{equation}
\hat{\psi} =e^{i \hat{\varphi} } \sqrt{\hat{\rho} }.
\label{2}
\end{equation}
The commutation
relation for the new operators is found to be
\begin{equation}
\hat{\rho}({\bf r}) \hat{\varphi}({\bf r}')-
\hat{\varphi}({\bf r}') \hat{\rho}({\bf r})=
 i \delta({\bf r}-{\bf r}').
\label{3}
\end{equation}
Using Eq.\ (\ref{2})  we rewrite the Hamiltonian (\ref{1}) as
\begin{equation}
H= \int d^2 r \bbox{ \{ } {\hbar^2\over 2 m}\{\sqrt{\hat{\rho}}
(\nabla \hat{\varphi })^2
\sqrt{\hat{\rho}} +(\nabla \sqrt{\hat{\rho}})^2  +
i [(\nabla \sqrt{\hat{\rho}}) (\nabla \hat{\varphi}) \sqrt{\hat{\rho}} -
\sqrt{\hat{\rho}} (\nabla \hat{\varphi}) (\nabla \sqrt{\hat{\rho}})]\}+
[u({\bf r})-\mu ] \hat{\rho } +{\gamma \over 2} \hat{\rho }^2\bbox{ \} }.
\label{4}
\end{equation}
The density operator can be decomposed as
\begin{equation}
\hat{\rho}({\bf r}) =\rho_0({\bf r})+\hat{\rho}_1({\bf r}),
\label{5}
\end{equation}
where $\rho_0({\bf r})$ is
determined by the condition that the expansion of Eq.\ (\ref{4})
into powers of $\hat{\rho}_1$
and $\nabla\hat{\varphi}$ does
not contain the linear terms. It yields the equation
\begin{equation}
u({\bf r})-\mu +\gamma \rho_0({\bf r})-
{\hbar^2\over 2 m}{1\over \sqrt{\rho_0({\bf r}) }}\nabla^2
\sqrt{\rho_0({\bf r})}=0.
\label{8}
\end{equation}
Eq.\ (\ref{8})  coincides with the Gross-Pitaevskii equation
for the time independent
part of the condensate wave function \cite{4}.
Therefore, the function $\rho_0(\bf r)$
can be understood as the density of the particles at $T=0$.
At low temperatures the fluctuations of the density and the current
are assumed to be small  and one can neglect the anharmonic terms.
The quadratic part of the Hamiltonian reads as
\begin{equation}
H_2 = \int d^2 r \bbox { \{ } {1\over 4} [ 2\sqrt{\rho_0} \hat{\varphi }
\hat{T}(2\sqrt{\rho_0}\hat{\varphi }) +
{\hat{\rho }_1 \over \sqrt{\rho_0}} \hat{G}
({\hat{\rho }_1 \over \sqrt{\rho_0}})]+
i{\hbar^2\over  4 m} \{ (\nabla \hat{\rho }_1) (\nabla \hat{\varphi }) -
(\nabla \hat{\varphi }) (\nabla \hat{\rho }_1) -{\nabla \rho _0\over \rho _0}
[ \hat{\rho }_1 (\nabla \hat{\varphi }) -
(\nabla \hat{\varphi })  \hat{\rho }_1]\}\bbox{\}},
\label{9}
\end{equation}
where
\begin{equation}
\hat{T}= -{\hbar^2\over 2 m} [ \nabla^2 -
{1\over\sqrt{\rho_0 }}(\nabla^2 \sqrt{\rho_0})],
\label{10}
\end{equation}
\begin{equation}
\hat{G} =\hat{T}+ 2 \gamma \rho _0.
\label{10a}
\end{equation}

Let us  rewrite the operators
$\hat{\varphi}$ and
$\hat{\rho}_1$
in the second quantization representation
in terms of creation ($\hat{b}_\nu^+$)
and annihilation ($\hat{b}_\nu $) operators of the elementary
excitations, which should also be bosons in the system of Bose
particles. It is convenient to write
$\hat{\varphi}$ and
$\hat{\rho}_1$ in the following form \cite{24}:
\begin{equation}
\hat{\rho }_1 = \sqrt{\rho _0}\sum_\nu  [ F_\nu  \hat{b}_\nu  +
F_\nu ^* \hat{b}^+_\nu ],
\label{11}
\end{equation}
\begin{equation}
\hat{\varphi } = {1\over 2 i \sqrt{\rho _0}}
\sum_\nu  [ \Theta_\nu  \hat{b}_\nu  -
\Theta_\nu ^* \hat{b}^+_\nu ].
\label{11a}
\end{equation}
The representation (\ref{11}), (\ref{11a})
should transform the Hamiltonian  (\ref{9})
to the diagonal form
\begin{equation}
H_2 = const + \sum_\nu  (E_\nu  + {1\over 2}) \hat{b}_\nu ^+ \hat{b}_\nu ,
\label{12}
\end{equation}
where $E_\nu $ is the energy of the excitation with the
quantum number $\nu $.
The Hamiltonian  (\ref{9}) is reduced to the form (\ref{12}), if
the functions $F_\nu (r)$, $\Theta_\nu (r)$ satisfy the equations
\begin{equation}
\hat{T}\Theta_\nu = E_\nu  F_\nu ,
\label{13}
\end{equation}
\begin{equation}
\hat{G}F_\nu = E_\nu  \Theta_\nu,
\label{13a}
\end{equation}
and are normalized by the condition
\begin{equation}
\int d^2 r \Theta_\nu ({\bf r}) F_\nu ^*({\bf r})=1.
\label{15}
\end{equation}
One can check that the operators $\hat{\rho }$, $\hat{\varphi }$ in the
form (\ref{11}), (\ref{11a})
with $F_\nu $, $\Theta_\nu $ determined by Eqs.\ (\ref{13}),
(\ref{13a}), and
(\ref{15}) satisfy the commutation relation (\ref{3}).

One can see that  under the substitution
$F_\nu =u_\nu  + v_\nu $,
$\Theta_\nu =u_\nu  - v_\nu $
Eqs.\ (\ref{13}), (\ref{13a})
coincide with the Bogoliubov-de Gennes equations \cite{4}.
We note also a similarity between our approach and the one developed by
Ho and Ma \cite{27}.

\section{COLLECTIVE MODE SPECTRUM IN THE ANTI-TRAP GEOMETRY}
\label{s3}

In this section we consider the special choice of the external
potential:
\begin{equation}
u(x,y) = \mu - u \cosh^2 {x \over l}.
\label{19}
\end{equation}
We assume the parameters of the potential satisfy the
inequality $u\gg \hbar^2/2 m l^2$. Then, one can neglect the
fourth term in the left-hand side
of Eq.\ (\ref{8}), that corresponds to the
Tomas-Fermi approximation.
At $T=0$ the density of the particles is equal to
\begin{equation}
\rho_0(x,y)={u\over \gamma } \cosh^2 { x\over  l}.
\label{20}
\end{equation}

To derive the spectrum of low-energy excitations ($E\ll u$) one can
omit the kinetic energy terms  $\hat{T}$ in the operator $\hat{G}$.
Then, Eqs.\ (\ref{13}), (\ref{13a})
yield the following differential equation for the
function $\Theta$
\begin{equation}
2 u \cosh^2 {x\over l}\ \hat{T} \Theta_\nu ({\bf r}) =
E_\nu ^2 \Theta_\nu ({\bf r}).
\label{23}
\end{equation}
The function $F$ is given by the expression
\begin{equation}
F_\nu ({\bf r}) = {E_\nu  \over 2 u} \cosh^{-2} {x\over l}\
\Theta_\nu ({\bf r}) .
\label{24}
\end{equation}

Taking into account  that the system is uniform in the $y$ direction
the function $\Theta_\nu $ can be presented in the form
\begin{equation}
\Theta_\nu (x,y)=\theta_\nu (x) e^{i k y},
\label{26}
\end{equation}
where $\theta_\nu (x)$ satisfies the equation
\begin{equation}
-{\hbar^2 u\over m} \cosh^2 {x\over l} \
[{d^2\theta_\nu \over d x^2} - (k^2+l^{-2})
\theta_\nu ] = E_\nu ^2 \theta_\nu.
\label{27}
\end{equation}

Eq.\ (\ref{27}) coincides with the Shroedinger equation for the one
particle problem in the potential
$v(x)=\cosh^{-2}(x/l)$.
Its solution given in Ref.\ \cite{28} has the form
\begin{equation}
\theta_\nu (x)= \cosh^{-\eta}{x\over l}\ w(\frac{1-\tanh \frac{x}{l}}{2}),
\label{28}
\end{equation}
where $\eta=\sqrt{1+(k l)^2}$, and the function $w(z)$
is governed by a hypergeometric differential equation
\begin{equation}
z(1-z) w'' +(\eta+1)(1-2 z) w' -(\eta-s)(\eta+s+1) w =0.
\label{29}
\end{equation}
In Eq.\ (\ref{29}) the parameter $s$ is
determined by the equation
$E^2_\nu =\omega_0^2 s (s+1)$, where
$\omega_0=\sqrt{{\hbar^2 u/ m l^2}}$.
To solve Eq.\ (\ref{29}) one should specify the boundary conditions.
We consider the system infinite in the $x$ direction. Since the
fluctuations of the current should remain finite,
the quantity $d (\theta(x)/\sqrt{\rho_0(x)})/d x$ cannot
be singular  at $|x|\to\infty$.
The solutions of Eq.\ (\ref{29}) satisfying such a condition are
the polynomials.
Eq.\ (\ref{29}) has the polynomial solutions at
$s=\eta+n$, where $n=0,1,2,\ldots$. Therefore, the energy of the
collective
modes are equal to
\begin{equation}
E_{n k}=\omega_0 \sqrt{[n+\sqrt{1+(k l)^2}][n+1+\sqrt{1+(k l)^2}]}.
\label{30}
\end{equation}
The exact expression for the eigenfunction $\theta_{n k}(x)$  reads as
\begin{equation}
\theta_{n k}(x)= C_{n k} \cosh^{-\eta}{x\over l}\
P_n^{\eta\eta}(\tanh{x\over l}),
\label{31}
\end{equation}
where $P_n^{\eta\eta}$ is the Jacobi polynomial.
The normalization factor $C_{n k}$ is calculated to be
\begin{equation}
C_{n k}= \left( {u_0\over l L_y E_{n k} }
{n! (2 n + 2 \eta +1) \Gamma(n+ 2 \eta +1)
\over 2^{2\eta+1}\Gamma^2(n+\eta+1)}\right)^{1/2}.
\label{32}
\end{equation}
In Eq.\ (\ref{32}) $\Gamma $ is the Gamma function, $L_y$, the
size of the system in the $y$ direction.

The answers obtained have two uncommon features. First,
the spectrum does not contain an acoustic branch.
Second, all the modes are localized near $x=0$. Such a peculiar behavior
is connected with  that the model  contains an unphysical assumption.
We fix the parameter $l$, while set an infinite size of the system
in $x$ direction. Thus, we allow the density of particles be unbounded at
$|x|\to\infty$.
The principal question has to be clarified is whether the results
of the model (further we call it the artificial model) can be
applied to real physical objects.

Let us consider two examples.
The first one is a 2D Bose cloud of a stripe shape.
We assume that there are two rigid walls at $x=\pm L_x$, and
the potential well obeys the internal structure given at
$|x|<L_x$ by Eq.\ (\ref{19}). Since the particles cannot flow
though the rigid walls, the gradient of the phase normal to the
wall should be zero at $|x|=\pm L_x$.
Therefore, the boundary conditions for the
function  $\theta(x)$  are
\begin{equation}
{d\over d x} \left( {\theta(x)\over \sqrt{\rho_0(x)}}\right)
\Big|_{x=\pm L_x} =0.
\label{33}
\end{equation}

\begin{center}
\begin{figure}
\centerline{\epsfig{figure=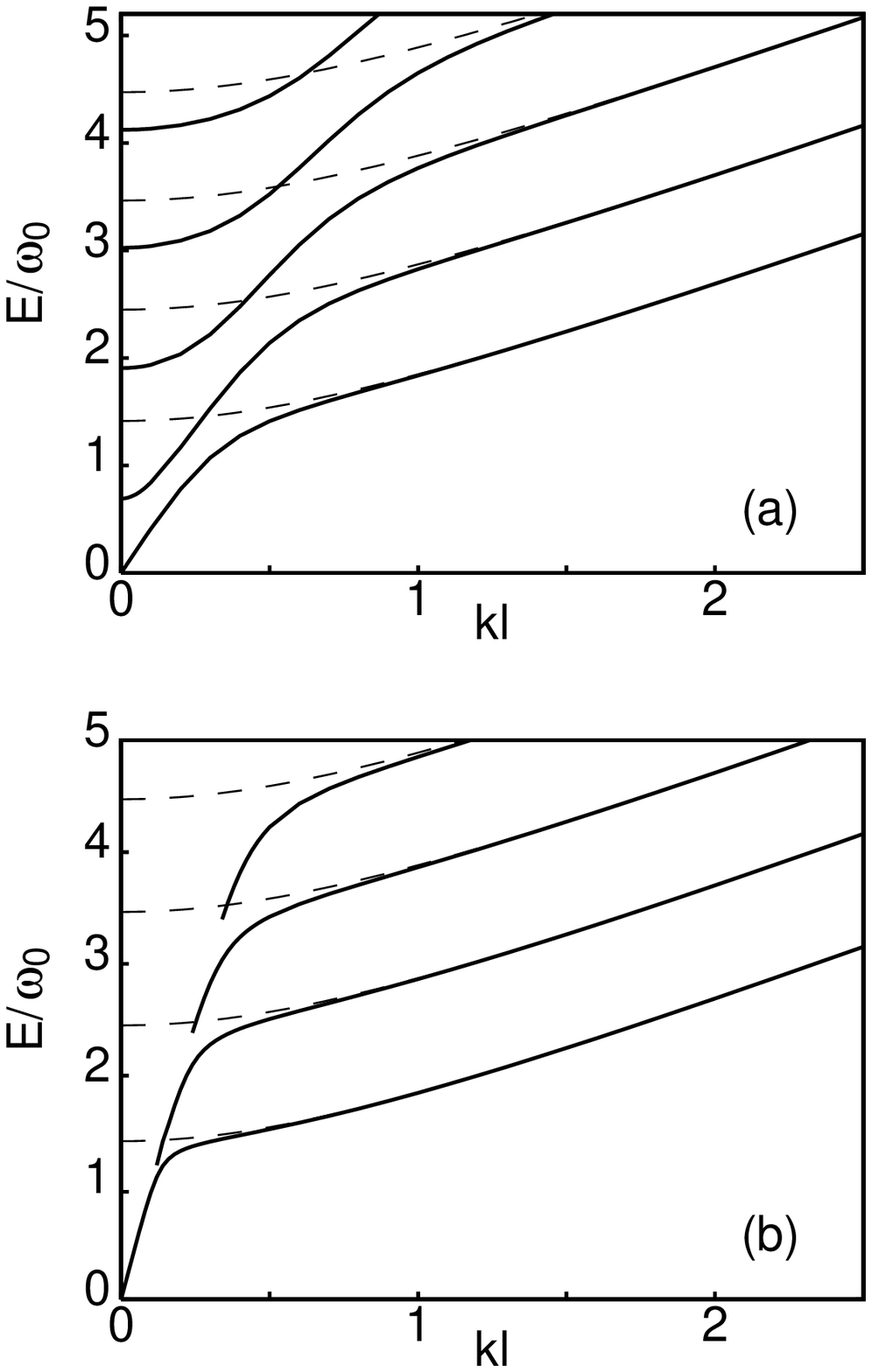,width=8cm}}
\vspace{0.5cm}
\caption{
Collective mode spectrum in the anti-trap potential for
the stripe (a) and the rare density
valley geometry (b).
Solid curves show the numerical results; dashed curves, the
result of Eq.\ (\ref{30}).
}
\label{f1}
\end{figure}
\end{center}

The general solution of Eq.\ (\ref{27}) can be written in the form
\begin{equation}
\theta(x) = \cosh {x\over l}\ [A_1 \varphi_1(x)+
A_2\varphi_2(x)],
\label{34}
\end{equation}
where
\begin{equation}
\varphi _1(x)=\cosh^{-\eta-1}{x\over l}\ \ _2F_1(\eta-s, \eta+s+1,
\eta+1,\frac{1-\tanh \frac{x}{l}}{2}),
\label{35}
\end{equation}
\begin{equation}
\varphi_2(x)=\cosh^{-\eta-1}{x\over l}\ (1-\tanh {x\over l})^{-\eta}
\  \ _2F_1(-s, s+1,1-\eta,\frac{1-\tanh \frac{x}{l}}{2}).
\label{35a}
\end{equation}
In Eqs.\ (\ref{35}), (\ref{35a})
$_2F_1(x,y,z,u)$ is the hypergeometric function.
We specify $1-\eta\ne -1, -2,\ldots$. (If $1-\eta$ is
negative integer, another form of the general solution should be used.)
Using Eqs.\ (\ref{33}), (\ref{34}) we get the dispersion
equation
\begin{equation}
\varphi_1'(L_x) \varphi_2'(-L_x) -
\varphi_1'(-L_x) \varphi_2'(L_x)=0.
\label{36}
\end{equation}
Its solutions can be obtained numerically. It is instructive
to analyze the case of $L_x\gg l$ (at $L_x\approx l$ the density variation
across the stripe is small).
The solution of Eq.\ (\ref{36}) at $L_x=3 l$ is shown in Fig.\ \ref{f1}(a).
For comparison, the spectrum given by the artificial
model (Eq.\ (\ref{30})) is also presented in this figure.
One can see that in the stripe system the lowest collective mode
is an acoustic one at small $k$. Its velocity is equal to
$\bar{c}= \sqrt{\gamma \bar{\rho}/ m}$, where
$\bar{\rho}$ is the average density of the particles.
There is a discrepancy in behavior of the two families
of the curves in Fig.\ \ref{f1}(a) only at
\begin{equation}
E>\hbar \bar{c} k.
\label{37}
\end{equation}
If the inequality opposite to  (\ref{37}) is satisfied, one can
neglect the influence of the boundaries. In this case the collective
mode spectrum is well approximated by Eq.\
(\ref{30}).

To consider another example
we specify the case of a Bose cloud
infinite in the $x$ direction with the density of the particles
given by Eq.\ (\ref{20}) at $|x|\le L_x$ and the uniform density
at  $|x|\ge L_x$.
There are localized and extended
collective modes in such a system.
The energies of the extended modes satisfy the inequality
(\ref{37}).
The localized mode solution has the form
$\theta(x)= A_3 \exp(\kappa x)$  at $x\le -L_x$,
$\theta(x)= A_4 \exp(-\kappa x)$ at $x\ge L_x$, where
$\kappa=\sqrt{k^2-E^2/\hbar^2 \bar{c}^2}$. At
$|x| \le L_x$ it is given by
Eq.\ (\ref{34}).
The boundary conditions are reduced to the continuity
of $\theta(x)$ and $d(\theta(x)/\sqrt{\rho_0(x)})/d x$ at
$x=\pm L_x$. The dispersion equation has the form
\begin{equation}
[\varphi_1'(-L_x)-\kappa \varphi_1(-L_x)]
[\varphi_2'(L_x)+\kappa \varphi_2(L_x)]   -
[\varphi_1'(L_x)+\kappa \varphi_1(L_x)]
[\varphi_2'(-L_x)- \kappa \varphi_2(-L_x)] =0.
\label{38}
\end{equation}
The solutions of Eq.\ (\ref{38}) at $L_x= 3 l$ are shown
in Fig.\ \ref{f1}(b).
This figure illustrates that
the analytical expression  (\ref{30}) yields a
satisfactory
approximation for the spectra of localized modes.
From the numerical analysis
we find that the coincidence
is better for larger ratio $L_x/l$.

Thus, our artificial model, while is not free from some artifacts,
seems quite useful for obtaining the analytical expressions for the spectra
and eigenfunctions of collective modes for real physical objects.

\section{COLLECTIVE MODE SPECTRUM IN THE LINEAR POTENTIAL}
\label{s4}

In this section we consider
the case of a linear external potential within the approach presented in
Sec.\ \ref{s3}.
This problem was studied preliminary in
Ref.\ \cite{24}.

The external potential has the form
\begin{equation}
u(x, y) =\mu -  \alpha x.
\label{39}
\end{equation}
Let us consider the low-energy excitations in the Tomas-Fermi regime.
In this regime the density of the particles
is equal to $\rho_0(x,y) = \alpha x/\gamma$.

The equation for $\theta(x)$ is reduced to
\begin{equation}
-{\hbar^2 \alpha \over  m } x
[{d^2\theta_\nu \over d x^2} + ({1\over 4 x^2}-k^2)
\theta_\nu ] = E^2_\nu  \theta_\nu .
\label{41}
\end{equation}
Substituting
\begin{equation}
\theta_\nu (x)=\sqrt{2|k|x}e^{-|k|x}w(2|k|x)
\label{42}
\end{equation}
into Eq.\ (\ref{41}), we obtain
the following equation
for the function $w(z)$:
\begin{equation}
z w'' +(1-z)w' - \beta  w =0
\label{43}
\end{equation}
with
\begin{equation}
\beta  = {1\over 2}\left( 1-{E^2_\nu  m\over  \hbar^2 |k| \alpha}\right) .
\label{44}
\end{equation}

At the beginning, we consider the artificial model.
We assume the system is semi-infinite and  the density is unbounded at
$x\to\infty$. We are interested in the solutions finite at
$x=0$ and $x\to\infty$. Such solutions exist if
$\beta =-n$ ($n=0,1,2,\ldots$).
Using Eq.\ (\ref{44}) we obtain the following analytical
expression for the energies:
\begin{equation}
E_{n k} =  \sqrt{{\hbar^2\alpha \over m} |k| (2 n + 1)}.
\label{45}
\end{equation}
The function $\theta_\nu (x)$ reads as
\begin{equation}
\theta_{n k}(x)=C_{n k} \sqrt{2|k|x}e^{-|k|x}L_n(2|k|x),
\label{46}
\end{equation}
where $L_n$ is the Laguerre polynomial. The normalization factor is
equal to
\begin{equation}
C_{n k} = \sqrt{{\alpha \over  L_y E_{n k}}}.
\label{47}
\end{equation}
As in the case of the anti-trap potential
the spectrum of the excitations does not contain an acoustic
branch. It is connected with the unbounded density at
$x\to\infty$ implied in the model.

\begin{center}
\begin{figure}
\centerline{\epsfig{figure=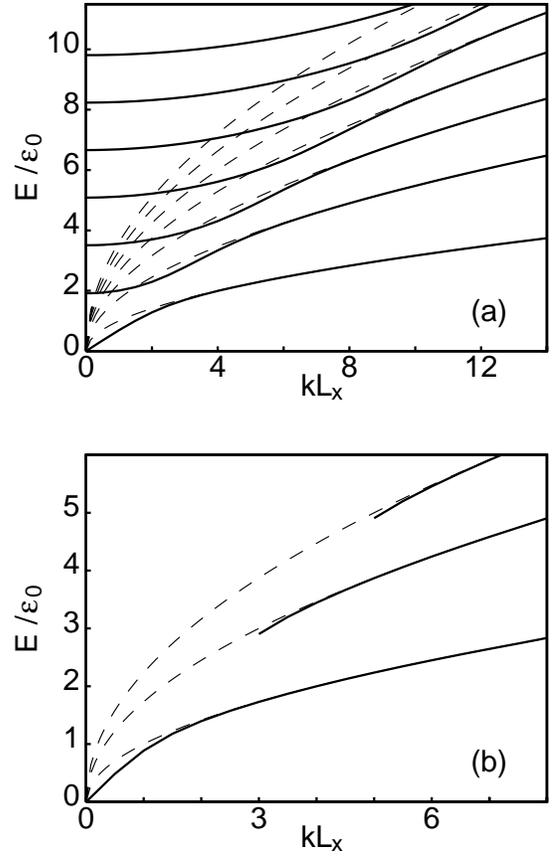,width=8cm}}
\vspace{0.5cm}
\caption{
Collective mode spectrum in the linear potential
for the stripe (a) and the semi-infinite system
(b).
Solid curves show the numerical results; dashed curves, the
result of Eq.\ (\ref{45}).
}
\label{f2}
\end{figure}
\end{center}

Let us switch to the real systems and consider the stripe
with a linear increase of the density across the stripe and
a rigid wall at $x=L_x$.
The solution of Eq.\ (\ref{41}) finite at
$x=0$ has the form
\begin{equation}
\theta(x)= A \sqrt{x}
e^{-|k|x}\
_1F_1(\beta ,1, 2|k| x),
\label{48}
\end{equation}
where $_1F_1(x,y,z)$ is the Kummer confluent hypergeometric function.
Using the boundary condition Eq.\ (\ref{33}) at $x=L_x$, we obtain
the following dispersion equation
\begin{equation}
_1F_1(\beta ,1, 2|k| L_x) - 2 \beta  \ _1F_1(1+\beta,2, 2|k| L_x)=0.
\label{49}
\end{equation}
Its solutions
are shown in Fig.\ \ref{f2}(a).
The spectrum (\ref{45}) is also presented in Fig.\ \ref{f2}.
In this figure the energy units $\epsilon_0=\sqrt{\hbar^2\alpha /m L_x}$ are
used.
One can see that
at $ k> E/ \hbar \bar{c}$ the spectrum is well approximated by the
analytical expression (\ref{45}), while at smaller $k$ the difference
is essential. In particular,
the lowest mode transforms to the
acoustic one.

To complete the picture we consider a semi-infinity system with
a linear increase of the density at
$0\le x\le L_x$ and the uniform density at  $x \ge L_x$.
We are interested in the edge collective excitations of
such a system.
The dispersion equation for the edge modes is found to be
\begin{equation}
(\kappa - |k|)\  _1F_1 (\beta , 1, 2|k| L_x) + 2 |k| \beta  \
_1F_1 (1+\beta,2, 2|k| L_x)=0.
\label{50}
\end{equation}
The solution of Eq.\ (\ref{50}) is shown in Fig.\ \ref{f2}(b).
One can see that the spectrum of the edge excitation is very close to the
dependence given by the analytical expression
(\ref{45}).

To conclude this section we discuss the conditions, when the low-energy
description is valid. The Tomas-Fermi approximation
is  correct at
\begin{equation}
\gamma \rho_0(x)\gg {\hbar^2\over 2 m x^2},
\label{40p}
\end{equation}
when the
fourth term in the left-hand side of Eq.\ (\ref{8}) can be neglected.
In two dimensions the parameter $\gamma$ is connected with
the radius of the interparticle interaction
$a$ by the relation \cite{Sc}
\begin{equation}
\gamma =-{4 \pi \hbar^2 \over m} {1\over \ln(\rho_0 a^2)}.
\label{22}
\end{equation}
Since Eq.\ (\ref{22}) contains only the factor $\ln(\rho_0 a^2)^{-1}$
as a small parameter, the condition  (\ref{40p}) is fulfilled
at $x>x_0$, where $x_0$ is of order of
the local average distance between the particles.

To omit the kinetic energy terms in the operator $\hat{G}$
we should
consider the excitations with the energies
\begin{equation}
E<\gamma \rho_0(x_0)=\alpha x_0.
\label{81}
\end{equation}
Since we neglect the solution of Eq.\ (\ref{43})
singular at $k x\to 0$,
the wave vector $k$ should satisfy the inequality
\begin{equation}
k x_0\ll 1.
\label{82}
\end{equation}
Using Eqs.\ (\ref{45}), (\ref{81}), (\ref{82}) we find the conditions on
the wave vectors and the energies
\begin{equation}
k\ll \left( {\alpha m\over \hbar^2}\right)^{1/3},
\label{83}
\end{equation}
\begin{equation}
E\ll E_m=\left( {\hbar^2 \alpha ^2\over m}\right)^{1/3}.
\label{83a}
\end{equation}
Eqs.\ (\ref{83}), (\ref{83a})
set the validity of the low-energy approximation.
At finite $L_x$ the  energies  have the scale of $\epsilon_0$. Therefore,
the description requires $\epsilon_0\ll E_m$, that yields the condition
on the size of the system in the $x$ direction
\begin{equation}
L_x \gg \left( {\hbar^2 \over \alpha m}\right)^{1/3}.
\label{84}
\end{equation}

\section{OFF-DIAGONAL LONG RANGE ORDER}
\label{s5}

It is of interest to consider the possibility of BEC
in 2D systems subjected to the anti-trap or linear potential.
Let us analyze the behavior of the one-particle density
matrix
\begin{equation}
W({\bf r},{\bf r}')=\langle \hat{\Psi}^+({\bf r})
\hat{\Psi}({\bf r}')\rangle.
\label{51}
\end{equation}
If the function $W({\bf r},{\bf r}')$ remains finite at
$|{\bf r}-{\bf r}'|\to\infty$,
the off-diagonal long range order is realized.

Thermal fluctuations destruct the macroscopic quantum coherence.
The principal mechanism of the destruction is connected with the
phase fluctuations. They are described by the operator
$\hat{\varphi}$. Neglecting a small contribution of the
density fluctuations we obtain the following expression for the
function $W({\bf r},{\bf r}')$:
\begin{equation}
W({\bf r},{\bf r}')\approx
\sqrt{\rho_0({\bf r})\rho_0({\bf r}')}
\exp({-\Phi({\bf r},{\bf r}')\over 2}),
\label{52}
\end{equation}
where
\begin{equation}
\Phi({\bf r},{\bf r}')=\langle[\hat{\varphi }
({\bf r})-\hat{\varphi }({\bf r}')]^2\rangle.
\label{53}
\end{equation}

To be more concrete, we analyze the dependence of
 $\Phi(r,r')$ on $y$, $y'$ at
$x=x'$. Taking into account Eq.\ (\ref{11}) we get
\begin{equation}
\Phi(x,y;x,y') =
{1\over \rho_0(x)}
\sum_{n=0}^\infty {L_y\over 2\pi }\int_{-\infty}^{+\infty}
d k  |\theta_{n k}(x)|^2
[n_B(E_{n k})+{1\over 2}][1-\cos k(y-y')],
\label{54}
\end{equation}
where $n_B(E)=[\exp(E/T)-1]^{-1}$.

Let us consider the case of the anti-trap potential.
Since the maximum of the phase fluctuations is located
at $x=0$ (see. Eq.\ (\ref{31})),
the phase coherence destruction begins at $x=0$.
Therefore, if the long range order survives at $x=0$, it
takes place in the whole system.

Using the analytical expressions
(\ref{30}), (\ref{31}) for
$E_{n k}$ and $\theta_{n k}(x)$
we obtain the following asymptotic result:
\begin{equation}
\Phi(0,y,0,y')={T\over 2 \pi T_0(0)  } \ln{l\over \xi_T(0)} +\Phi_{\rm zp},
\label{55}
\end{equation}
where $T_0(x)=\hbar^2\rho_0(x)/2 m$,
$\xi_T(x)=\sqrt{\hbar^2\gamma \rho_0(x)/m T^2}$.
In Eq.\ (\ref{55}) the inequalities
$l\gg \xi_T(0)$, $|y-y'|\gg l$ are implied.
The term $\Phi_{\rm zp}$
describes a contribution of the zero-point fluctuations. This term
determines the density of the condensate $n_0(x)$ at $T=0$.
The temperature dependent part in the right-hand side
of Eq.\ (\ref{55})
does not depend on $|y-y'|$. It means that the condensate survives
at nonzero temperatures.
According to Eq.\ (\ref{55}) the condensate density at $T\ne 0$
is connected with $n_0(0)$ by the expression
\begin{equation}
n_T(0) =n_0(0)\Big({\xi_T(0)\over l}\Big)^{\frac{T}{4\pi T_0(0)}}.
\label{56}
\end{equation}
The result (\ref{56}) is obtained within the artificial model.
Nevertheless, one can apply it to the real objects, if in Eq.\ (\ref{54})
the contribution of the modes with the energies (\ref{37}) is
negligible. For example,
such a situation is realized for
a Bose cloud of the shape of a
square $L\times L$  with
$L/l\gg 1$. Then, the answer
(\ref{56}) yields a satisfactory approximation for the density of the
condensate.  For the square geometry
the total number of the particles
$N$ is connected with the parameter of the model
by the relation
\begin{equation}
{N\over L^2}={u\over 2 \gamma }(1+ {l\over L} \sinh { L\over l}).
\label{57}
\end{equation}
It is natural to define the thermodynamic limit as
$N\to\infty$, $L\to\infty$ at $N/L^2=const$.
From Eq.\ (\ref{57}) we also obtain
$L/l=const$.
Therefore,
in the thermodynamic
limit
the parameter $l$ depends on
the total number of the particles ($l\sim N^{1/2}$).
From Eq.\ (\ref{56}) we find that the density of the condensate decreases
by the law $n_T(0)\sim N^{-b}$ ($b>0$)
under increase of $N$.
It indicates the absence of true BEC in the thermodynamic limit.
At finite $N$, as it follows from Eq.\ (\ref{56}), the BEC regime
can be realized.

Similar calculations for the linear potential yield
\begin{equation}
n_T(x) =n_0(x)\Big({\xi_T(x)\over x}\Big)^
{\frac{T}{4\pi T_0(x)}}.
\label{58}
\end{equation}
In Eq.\ (\ref{58}) the inequalities $x\gg\xi_T(x)$,
$|y-y'|\gg x$ are implied.  For a Bose cloud of the shape of a
rectangle $L_x\times\L_y$
the thermodynamic limit
corresponds to $N\to\infty$, $L_x\to\infty$, and $L_y\to \infty$
at $N/(L_x L_y)=const$ and $L_x/L_y=const$.
To analyze the density of the condensate at the same
local density of the particles
we should also fix the ratio $x/L_x$ in Eq.\ (\ref{58}).
Then again $n_T(x)$ approaches to zero at $N\to\infty$ and
there is no BEC in the thermodynamic limit.

\section{CONCLUSIONS}
\label{s6}

We  have considered the low-energy collective excitations
of the interacting inhomogeneous Bose gas in two-dimensions.
We specify the two types of the external potential: the anti-trap
one, in which the rare density valley is formed, and the linear potential,
that models the simplest case of the edge geometry. We  show that
for the configurations considered an approximate analytical solution
for the energies and the eigenvectors of the collective modes
can be obtained.
To derive the analytical solution we consider the artificial model,
in which an unbounded density of the particles is allowed.
Physically, such an approximation means that we suppress the
fluctuations in the high-density regions. Our method yields the modes
localized at the rare density valley or at the edge. Their spectra does
not contain an acoustic branch. The modes lost in our approach can be
obtained from the numerical solution of the eigenvalue problem. In this case
we should specify the high-density profile. We present the results of the
numerical analysis and compare them with the analytical solutions. We
find that the spectra obtained numerically are well approximated by the
analytical expressions in case of $E<\hbar k\sqrt{\gamma \bar{\rho}/m}$,
where $E$ is the energy of the excitation, $k$, the wave vector along the
valley or the edge, $\bar{\rho}$, the total average density.
It is important to note that in this case the spectrum of the
collective modes
is determined completely by the parameters of the external potential and
does not depend on the interaction between bosons. Collective modes of
the trapped Bose gas in the Tomas-Fermi regime demonstrate the same
behavior \cite{14}.

The possibility of BEC in the systems under consideration has been
investigated. The high density regions affect significantly on the
character of the thermal fluctuations in the low density areas.
If the total number of the particles is finite,
the BEC regime is realized at nonzero temperatures. This conclusion is in
agreement with the results of Ref.\ \cite{10}, where the BEC regime of
a 2D Bose gas in the trapping potential has been described.
The proper formulation of the thermodynamic limit for the systems
considered shows that in this limit
the true BEC does not occur, as it should follow
from the general theorems \cite{9,29,30}.

\acknowledgments

The authors would like to thank G.\ V.\ Shlyapnikov for helpful discussions.
This work was supported  by the INTAS Grant No.\ 97-0972.

\eject
\begin{multicols}{2}

\end{multicols}

\end{document}